\documentstyle[preprint,aps,prl]{revtex}

\begin{document}

\title{Exponential Divergence and Long Time Relaxation in Chaotic Quantum
Dynamics}

\author{Arjendu K. Pattanayak and Paul Brumer} 

\address{ Chemical Physics Theory Group \\
University of Toronto \\ Toronto, Canada M5S 1A1 }
\date{April 10th, 1996}

\maketitle
\draft
\begin{abstract}
Phase space representations of the dynamics of the quantal and classical cat 
map are used to explore quantum--classical correspondence in a K-system: 
as $\hbar \to 0$, the classical chaotic behavior is shown to emerge
smoothly and exactly. The quantum dynamics near the classical limit displays 
both exponential separation of adjacent distributions and long time 
relaxation, two characteristic features of classical chaotic motion.
\end{abstract}

Understanding the correspondence principle for bound conservative
systems which are classically chaotic presents a serious challenge
\cite{ford1}. The classical system is characterized by positive K-entropy 
which is reflected, for example, in the exponential divergence of initially 
nearby trajectories and in the relaxation\cite{relax} of distributions in 
the long-term limit. By contrast, the quantum dynamics is not even ergodic, 
and displays long time recurrences.
Current views on correspondence in conservative chaotic systems range
from simple belief in the correspondence principle, to reliance 
on decoherence induced by coupling to an environment\cite{zurek}, to calls 
for the overthrow of quantum mechanics due to its purported inability to
display chaos in the correspondence limit\cite{ford1}. Resolving this
controversy requires studies of quantum dynamics close to the classical limit,
a task thus far made impossible by numerical difficulties as $\hbar
\rightarrow 0$.

In this Letter we demonstrate, using numerical results based on an
exact quantum propagator\cite{jw}, the approach to the classical limit of the
chaotic Arnold cat\cite{arnold}. The nonsingular character of our phase space
propagator as $\hbar \rightarrow 0$ allows direct insight into
correspondence in this bounded chaotic system. In particular, we show that 
quantum mechanics does display chaos as $\hbar \rightarrow 0$, including 
exponential divergence of initially nearby distributions and long time 
relaxation. The agreement between quantum and classical dynamics extends over 
longer times as $\hbar$ decreases\cite{logtime}.

Classical-quantum correspondence is best examined \cite{jaffe}
in the Liouville picture classically and in some phase
space representation of the density matrix (e.g. the Wigner
representation\cite{groot}) quantum mechanically. This approach provides an
overlap of objects and concepts which does not exist if one attempts
to utilize, for example, classical trajectories and
quantum wavefunctions. The advantages of this approach are evident in
the results below where operationally chaotic behaviour is demonstrated
in quantum mechanics.

The classical cat map is a K-system which corresponds to the classical 
dynamics of a kicked oscillator with Hamiltonian\cite{ford} 
\begin{equation}
H=p^2/2\mu +\epsilon x^2/2\sum_{s=-\infty}^{\infty}\delta(s-t/T).
\label{fh}
\end{equation}
restricted to a torus $0\leq x < a$, $0\leq p < b$. 
The proper quantization of this system and its approach to the classical
limit has been extensively considered\cite{ford,berry,jw}. Early 
quantizations \cite{berry} of this system were unsettling insofar as: 
(a) only a restricted class of classical tori (those with $ab=h N$, $N$ 
integer) could be quantized; (b) the solutions violated the uncertainty 
principle and (c)  the quantum propagator thus obtained did not reduce to 
the classically chaotic cat map in the $\hbar \rightarrow 0$ limit. 

A recently introduced\cite{jw} phase space quantization
eliminated these problems and yielded an analytic
quantum propagator. Specifically, the one time-step quantum propagation of 
the Wigner--Weyl representation of the quantal density operator
$\rho^Q(x,p;t)$ in phase space $(x,p)$ is given by
\begin{equation}
\rho^Q(x,p;T)=\frac{1}{ab}\sum_{n,m=-\infty}^{\infty}
\rho^Q_{n,m}(T)f_{n,m}
\label{form}
\end{equation}
where
\begin{equation}
\rho_{n,m}^{Q}(T)=\sum_{k,l=-\infty}^{\infty}G(n,m;k,l)\rho_{k,l}^
{Q}(0).
\label{ItoT}
\end{equation}
The propagator $G$ of the Fourier coefficients is given by
\begin{eqnarray}
&&G(n,m;k,l)=e^{
i\pi\alpha(kl-nm)}\int_0^1d\nu\int_0^1d\nu^{\prime}e^{-i\pi\xi[(\nu+\alpha
n){\rm mod}~1]^2/\alpha}e^{i\pi\xi\nu^2/\alpha}\cdot\nonumber \\
&&e^{2\pi i(l-m)\nu}
e^{-i\pi\eta[(\nu^{\prime}+\alpha
l){\rm mod}~1]^2/\alpha}e^{i\pi\eta\nu^{\prime 2}/\alpha}e^{2\pi
i(k-n)\nu^{\prime}}
\label{gprop}
\end{eqnarray}
where $\eta=Tb/\mu a\in\bf{Z}$ (the set of integers) and 
$\xi=-\epsilon Ta/b\in\bf{Z}$ obtain in scaling the original Hamiltonian to 
the unit torus, $\alpha=h/ab$ acts as a dimensionless form of Planck's 
constant for this problem and $f_{n,m}(p,x)=\exp\{2\pi i(n p/a +m x/b)\}$ is 
the Fourier expansion basis. 
The Arnold cat map\cite{arnold} corresponds to the choice $\eta=\xi=1$.
Equation (\ref{gprop}) can be integrated exactly\cite{jwdiss}
to provide an analytic propagator for all $\hbar$. 
This result contrasts with the classical propagator given by Eq.(\ref{form})
but with $\rho^Q_{n,m} (T)$ replaced by the classical form
\begin{eqnarray}
\rho_{n,m}(T)&=& \sum_{k,l = -\infty}^\infty G_c(n,m;k,l) \rho_{k,l}(0)
\nonumber \\ G_c(n,m;k,l)&=& \delta_{(k,l), \phi^T\cdot(n,m)}.
\label{gc}
\end{eqnarray}
Here $\phi^T\cdot(n,m)$ is the vector resulting from premultiplying the the 
vector $(n,m)$ with the transpose $\phi^T$ of the matrix governing the 
point dynamics:
\begin{equation}
\left( \begin{array}{c}
\tilde{x}_{n+1} \\
\tilde{p}_{n+1}
\end{array} \right)
=\phi\left( \begin{array}{c}
\tilde{x}_{n} \\
\tilde{p}_{n}
\end{array} \right){\rm mod}~1
=
\left( \begin{array}{cc}
1&\eta \\
\xi &1+\eta\xi
\end{array} \right)
\left( \begin{array}{c}
\tilde{x}_n \\
\tilde{p}_n
\end{array} \right)~{\rm mod}~1,
\label{cmatrix}
\end{equation}
with $\tilde{x}=x/a, \tilde{p}=p/b$.

Note that, by contrast with the classical case, the quantum propagator 
$G$ mixes contributions from all $k,l$ components to produce each $n,m$.
It shares with the classical propagator a fixed point at $(n,m)=(0,0)$.
However, it also has an infinite number of additional fixed points for 
rational $\alpha=u/v$; these lie in the quantal Fourier space at all 
$(wv,zv)$ with  $u,v,w,z \in {\bf Z}$. These fixed points affect
the quantal evolution of a distribution in the following ways: (a) near
the classical limit, ($\alpha \ll 1$ or $v \gg u \ge 1$), these fixed
points are far out in the Fourier space, and hence affect 
structure at extremely small scales. Thus, it is in the highly oscillatory 
regime (whether the structure is prepared initially or acquired during 
the evolution) that the dynamics will exhibit quantal evolution distinct 
from the classical dynamics; (b) at larger $ \alpha$, these fixed
points are both more dense in the Fourier space and exist at smaller
values of $(n,m)$, indicating a far more substantial
deviation from the classical evolution; (c) even though there are no
such demonstrable fixed points for irrational $\alpha$, the propagator is a 
smooth function of $\alpha$;
the qualitative conclusions from these arguments should therefore extend to
all $\alpha$. 

The evolution of distributions which are initially of the Gaussian form
\begin{equation}
N_\lambda \exp [-(x- \lambda x_0)^2/ \lambda^2 \sigma_x^2] \exp 
[-(p- \lambda p_0)^2/ \lambda^2 \sigma_p^2]
\label{lambda}
\end{equation}
is displayed below.
Here $N_\lambda$ is a normalization factor, $(\lambda x_0, \lambda
p_0)$ specifies the location of the Gaussian of width $(\lambda \sigma_x,
\lambda \sigma_p)$ in a phase space of dimension $\lambda^2ab$. The
variable scaling factor $\lambda$ can be used to approach the classical limit. 
That is, as $\lambda$ increases the volume of phase space $\lambda^2ab$ 
increases and $\alpha=h/\lambda^2ab$, the effective Planck's constant of the
problem, decreases. Hence, by increasing $\lambda$ we approach the classical 
limit while preserving the ratio ($\lambda^2 \sigma_x\sigma_p/\lambda^2ab$) 
of the volume of the initial distribution to the volume of phase space. 
We note that we work in units such that $h=1$.

Figures 1--3 show gray-scale contour maps of the evolution\cite{footnote}
of the Wigner function for $\alpha=10^{-1}, 10^{-2}$ and $ 10^{-5}$.
To simplify the presentation we report the absolute value of the Wigner
representation of the density. The Wigner--Weyl function for 
$\alpha=10^{-5}$ is always positive definite and numerically {\em identical} 
to the classical Liouville density over the indicated time scale.
We note that the classical dynamics is scale--invariant; hence the classical 
phase--space figures are independent of $\lambda$ and are the same for all 
three cases and identical to the $\alpha=10^{-5}$ case.
The analytic results\cite{jw} ensure that the $\alpha \to 0$
limit is indeed classical mechanics but these computations allow a
direct examination of the approach to this limit.

For $\alpha=0.1$ (Fig.~1) the system is seen to be in the deep quantum
regime. The propagating wavepacket shows only the barest resemblance to the 
classical result and interference structures dominate. There is no relaxation 
of the dynamics over the indicated timescale and the expectation value of 
operators (not shown) oscillate with large amplitude. 
Although not evident in this figure, the Wigner
distribution has many positive and negative regions. For $\alpha=0.01$ 
(Fig.~2) the classical density striations begin to appear, although 
still accompanied by quantum effects. Various parts of the distributions 
appear to interfere with one another giving interference fringes. Once again 
relaxation is not evident after eight time steps, although a large coarse 
graining will result in a uniform phase space distribution. 
Separate calculations show that the expectation value of operators in this 
$\alpha$ regime display small fluctuations about the equilibrium value. 
Finally, as noted above, the $\alpha=10^{-5}$ result agrees exactly with the
figures from the classical evolution: The distribution has relaxed\cite{relax}
even over the short time--scale shown. The approach to classical mechanics is 
clearly smooth and devoid of singularities.

Relaxation is but one characteristic of chaotic motion, evident at long times. 
Classical chaos, however, is typically manifested in the short time 
instability associated with exponential divergence of adjacent trajectories. 
Standard arguments\cite{wyatt} imply that such exponential separation is not 
possible in quantum mechanics. Here we demonstrate that this is not the case; 
we have obtained similar exponential divergence for compact phase space 
{\em distributions}, both classically and quantally close to the classical 
limit. In the absence of point trajectories and an appropriate tangent space, 
we examine the time dependence of the distance $\Delta$ between two nearby, 
initially well localized Gaussians\cite{param}; the results are shown in 
Fig.~4. Also shown for comparison is a straight line with slope dictated by 
the K-entropy of the classical cat map; since the system is uniformly 
hyperbolic, this is exactly equal to the Lyapunov exponent of the underlying 
point dynamics. The distance in phase space between the two distributions is 
determined by the centroids: 
$\Delta={[({\langle x \rangle}_1 - {\langle x \rangle}_2)^2+({\langle p 
\rangle}_1 - {{\langle p \rangle}_2)^2]}}^{1/2}$, where 
${\langle \rangle}_i$ denotes the average over the $i^{th}$ distribution 
($i=1,2$). The quantity $\Delta(t)$ was evaluated both classically and
quantum mechanically for various values of $a,b$.

The time dependence of $\Delta$ was found to be essentially identical
in classically and quantally propagated distributions over times up to eight 
steps, for the $a,b$ (and hence $\alpha$) in the range shown.
At early times, dependent on $\alpha$, $\Delta$ increases essentially 
exponentially [see Fig.~4], with increasing adherence to the classical 
K-entropy line as $\alpha$ is reduced.
At times longer than that shown in Fig.~4 $\Delta$ tends to decrease in a 
generally similar way quantum mechanically and classically, a consequence of 
saturation\cite{saturate} and relaxation in the finite size phase space.
When $\alpha$ is increased, since there is no relaxation, $\Delta$ continues 
to oscillate at larger times quantum mechanically, whereas the classical
dynamics relaxes. The combination of short--time exponential divergence and 
long--time relaxation demonstrates the loss of information characteristic of 
chaos for both the classical and near--classical quantal dynamics. 

Finally, we note that the availability of an analytic quantum propagator for a 
K-system allows a study of the expected time--dependence of classical-quantum
agreement as a function of $\hbar$. Preliminary results suggest that 
increasing the value of $\alpha$ does, in general, decrease the time-scale 
over which quantum and classical dynamics agree. However, the extent of 
agreement also depends strongly on various other factors, including the 
nature of the initial distribution.
The variation within the quantum--classical deviations for different initial 
conditions at a given value of $\alpha$ and $t$ spans many orders of 
magnitude. Further studies on this issue await our application of our general
quantization approach to classical maps which have smaller K-entropy than the
classical cat map. That is, the large value of the K-entropy for the cat map
forces relaxation of the dynamics over a rather short time. 
Further, it also rapidly saturates the exponential divergence [Fig.~4] which 
could otherwise extend to significantly longer times.

In summary, although the literature suggests that demonstrating 
quantum--classical correspondence becomes increasingly difficult (if not
impossible \cite{ford1,ford}) with increasing classical chaos, we have been 
able to show this correspondence in a classical K-system.  Secondly, we 
have shown how correspondence emerges smoothly from the quantal behavior
as $\alpha \to 0$.  These results make clear that the dynamics 
of distributions in the non-ergodic quantal cat map display the essential 
attributes of the classical K-system as the system approaches the classical 
limit. They constitute the first demonstration of the emergence of 
characteristically classically chaotic behavior from quantum mechanically 
non-ergodic behavior as one approaches this limit.

{\bf Acknowledgment:}
This work was supported by the Natural Sciences and Engineering Research
Council of Canada.

\begin{figure}[htbp]
\caption{Snapshots of $|\rho^Q(x,p,t)|$.Densities are represented by five 
shadings of grey denoting five evenly spaced probabilites between 
$7 \times 10^{-5}$ to $1 \times 10^{-6}$.} (a) $t =0$,(b) $t =2T$,
(c) $t =4T$,(d) $t =8T$; $\alpha = 0.1~(\lambda a=\lambda b=3.162$). 

\end{figure}

\begin{figure}[htbp]
\caption{As in Figure 1 but with $\alpha = 10^{-2} ~(\lambda a= \lambda b =$
10).}
\end{figure}

\begin{figure}[htbp]
\caption{As in Figure 1 but with $\alpha = 10^{-5} ~(\lambda a = \lambda b =
316.2)$.}
\end{figure}

\begin{figure}[htbp]
\caption{Short--time exponential separation of centroids for
$ a=b=50.0 \;(\alpha=4 \times 10^{-4}) $ (dashed) ; $a=b=28.79 \; (\alpha = 1.207 
\times 10^{-3}) \ $ (solid) ; $a=b=16.57 \; (\alpha =3.64 \times 10^{-2})
$ (dotted). See text for other parameters. Similar results are obtained by 
propagating the same distributions classically. The straight line with a 
slope of $0.9624$ shows classical exponential separation of adjacent point 
trajectories.}
\end{figure}


\begin{thebibliography}{99}
\bibitem{relax}The term relaxation  denotes the evolution of all initial 
distribution to the same stationary final distribution. 
In the case of the cat map, this is the constant distribution in phase space.
\bibitem{footnote} Wavepackets were initialized in phase space,
Fourier transformed and then propagated via Eq.~(\ref{gc}) or 
Eq.~(\ref{gprop}) using a $512 \times 512$ grid; various accuracy checks were 
performed using a $1024 \times 1024$ sized grid. The classical Fourier 
trajectories are hyperbolic; thus the Fourier components leave any finite 
grid rapidly. This is the major source of computational error classically 
and in the quantal $\alpha \to 0$ regime. Dynamics in the deep quantal 
regime ($\alpha \to 1$) do not exhibit any Fourier space hyperbolicity, 
thus avoiding the associated accuracy problems.
\bibitem{param} One Gaussian has parameters 
$\lambda=1,x_0 =0.00712,p_0=0.00904,\sigma_x= 0.4153,\sigma_p=0.3841$; 
the second Gaussian is identical except for $x_0=0.00702$. The values of 
$\alpha$ for this set of computations were chosen arbitrarily in the 
near--classical regime. The parameters for the Gaussians were also chosen 
arbitrarily, constrained solely to be initially well--localized while
satisfying the uncertainty principle.
\bibitem{saturate}The distance between two centroids, or between two point 
trajectories, is bounded due to the finite phase--space volume. Reaching this 
limit is what we term saturation.
\end{thebibliography}
\end{document}